\documentclass[conference]{IEEEtran}
\IEEEoverridecommandlockouts
\usepackage{cite}
\usepackage{amsmath,amssymb,amsfonts}
\usepackage{algorithmic}
\usepackage{graphicx}
\usepackage{textcomp}
\usepackage{xcolor}

\def\BibTeX{{\rm B\kern-.05em{\sc i\kern-.025em b}\kern-.08em
    T\kern-.1667em\lower.7ex\hbox{E}\kern-.125emX}}

\begin{document}

\title{Detection of Sleep Oxygen Desaturations from Electroencephalogram Signals\\

\thanks{NCH Sleep DataBank was supported by the National Institute of Biomedical Imaging
and Bioengineering of the National Institutes of Health under Award Number
R01EB025018. The National Sleep Research Resource was supported by the U.S.
National Institutes of Health, National Heart Lung and Blood Institute (R24
HL114473, 75N92019R002).}}

\author{
\IEEEauthorblockN{Shashank Manjunath}
\IEEEauthorblockA{\textit{Khoury College of Computer Sciences} \\
\textit{Northeastern University} \\
Boston, USA \\
\texttt{manjunath.sh@northeastern.edu}}

\and

\IEEEauthorblockN{Aarti Sathyanarayana}
\IEEEauthorblockA{\textit{Bouv\'e College of Health Sciences} \\
  \textit{Northeastern University} \\
Boston, USA \\
\texttt{a.sathyanarayana@northeastern.edu}}
}

\maketitle

\begin{abstract}
In this work, we leverage machine learning techniques to identify potential
biomarkers of oxygen desaturation during sleep exclusively from
electroencephalogram (EEG) signals in pediatric patients with sleep apnea.
Development of a machine learning technique which can successfully identify EEG
signals from patients with sleep apnea as well as identify latent EEG signals
which come from subjects who experience oxygen desaturations but do not
themselves occur during oxygen desaturation events would provide a strong step
towards developing a brain-based biomarker for sleep apnea in order to aid with
easier diagnosis of this disease. We leverage a large corpus of data, and show
that machine learning enables us to classify EEG signals as occurring during
oxygen desaturations or not occurring during oxygen desaturations with an
average 66.8\% balanced accuracy. We furthermore investigate the ability of
machine learning models to identify subjects who experience oxygen desaturations
from EEG data that does not occur during oxygen desaturations. We conclude that
there is a potential biomarker for oxygen desaturation in EEG data.
\end{abstract}


\section{Introduction}

Sleep apnea is a disease characterized by repeated stoppages of breathing during
sleep and is further associated with blood oxygen
desaturations~\cite{karhuLongerDeeperDesaturations2021}. In this work, we aim to
identify potential biomarkers of sleep apnea by assessing whether quantitative
measures of brain electrodynamics correlate with blood oxygen desaturation.
Recent research indicates the promise of signal processing technology to
automatically identify sleep apnea in
patients~\cite{manjunathTopologicalDataAnalysis2023a}. These desaturations can
be caused by physical obstructions in the case of obstructive sleep apnea, or by
lack of respiratory effort caused by central nervous system effects in the case
of central sleep apnea. However, while the underlying causes are varied, sleep
apnea is diagnosed by overnight sleep studies, called polysomnograms (PSGs),
which identify repeated apnea episodes during
sleep~\cite{prisantObstructiveSleepApnea2007}.

Since sleep apnea is caused by nighttime effects, but leads to symptoms when
awake, we hypothesize that latent central nervous system effects may be
distinguished by automated algorithms in order to identify oxygen desaturation
and related effects~\cite{carneiro-barreraWeightLossLifestyle2019}. To this end,
we leverage electroencephalogram (EEG) signals collected during sleep, and
develop machine learning techniques to answer the following questions:

\begin{enumerate}
  \item Can machine learning models identify EEG signals which occur during
    sleep oxygen desaturations? \label{rq1}
  \item Can machine learning models identify a latent marker of oxygen
    desaturation from EEG signals? \label{rq2}
\end{enumerate}

Development of a machine learning technique which can successfully identify EEG
signals from patients with sleep apnea as well as identify latent EEG signals
which come from subjects who experience oxygen desaturations but do not
themselves contain oxygen desaturation events would provide a strong step
towards developing a brain-based biomarker for sleep apnea in order to aid with
easier diagnosis of this disease.

\section{Methods}

\subsection{Data}\label{subsec:data}

We leverage the Nationwide Children's Hospital (NCH) Sleep DataBank (SDB) for
this work. This dataset contains PSG studies from 3,984 pediatric
patients~\cite{zhangNationalSleepResearch2018b,leeLargeCollectionRealworld2022b}.
Each PSG study contains a significant number of sensors, including EEG and pulse
oximetry. The dataset also contains technician labeling of
sleep events, including sleep stage (NREM1, NREM2, NREM3, and REM) as well as
oxygen desaturation and apnea events. We include data collected
at 256Hz from the F3-M2, F4-M1, C3-M2, C4-M1, O1-M2, O2-M1, and Cz-O1
differential channels. We perform subject selection for each sleep stage
independently. To select which patients experience oxygen desaturations in a
given sleep stage, we include patients that have an oxygen desaturation event
labeled during that sleep stage, have at least one ``apnea'' event labeled at
any point during their sleep study, and whose blood oxygenation (SpO2) drops
below 90\% during the oxygen desaturation event. These EEG epochs are labeled as
containing desaturations. We call this group the ``desaturated'' group. To
select which patients do not experience oxygen desaturations in a given sleep
stage, we include patients that do not have any ``apnea'' events labeled at any
point during their PSG study and  who do not have any oxygen desaturations below
95\% blood oxygenation during their PSG study at any sleep stage. We call this
group the ``undesaturated'' group.

\subsection{Signal Preprocessing}\label{subsec:signal_preproc}

We first eliminate 60 Hz and 120 Hz power noise using a 3rd order Butterworth
bandstop filter. Each labeled EEG epoch is 30 seconds long. In order to allow
for both robust feature extraction from each EEG signal, as well as maintain
sufficient time and frequency resolution for high-frequency oscillations
contained in each EEG signal, we leverage the discrete-time short-time Fourier
transform (STFT)~\cite{grochenigFoundationsTimeFrequencyAnalysis2001a}. We use a
Hann window as our window function. We implement the STFT and preprocessing
pipeline in Python using the scipy
package~\cite{virtanenSciPyFundamentalAlgorithms2020}.

\subsection{Age/Gender Matching}

Our dataset contains a large number of subjects, but we have an unequal
distribution of patients of each gender and age range. Training a model directly
on this unequal distribution of subjects would lead to a biased model and biased
results. In order to avoid this situation, we develop two age and gender
matching schemes for our data -- one to maximize the amount of subjects in our
model training and testing pipeline, and another to ensure equality in the
number of subjects in each age/gender group. We first segment our data into the
following age groups based on pediatric development: 0--2 , 2--5, 5--8, 8--12,
and 12--18. We further subdivide these groups into male/female to create 10
age/gender groups. We describe the subject counts in each group in
Table~\ref{table:age_gender_groups}.

\begin{table}[t]
\caption{Desaturated and Undesaturated Subject Counts for Age/Gender Groups}
\begin{center}
\begin{tabular}{|c || c|c|c|c || c|c|c|c|}
  \hline
  \textbf{Group} & \multicolumn{4}{|c||}{\textbf{Desaturated}} & \multicolumn{4}{|c|}{\textbf{Undesaturated}} \\
  \cline{2-9}
                            & N1 & N2 & N3 & R & N1 & N2 & N3 & R \\
  \hline
  \hline
  0--2 Female   & 69  & 95  & 88  & 112 & 1  & 1  & 1  & 1  \\
  \hline
  0--2 Male     & 115 & 159 & 129 & 178 & 2  & 2  & 2  & 2  \\
  \hline
  2--5 Female   & 129 & 157 & 109 & 178 & 5  & 5  & 5  & 4  \\
  \hline
  2--5 Male     & 160 & 182 & 141 & 197 & 2  & 3  & 3  & 3  \\
  \hline
  5--8 Female   & 75  & 91  & 63  & 99  & 6  & 7  & 6  & 7  \\
  \hline
  5--8 Male     & 101 & 124 & 94  & 128 & 10 & 10 & 10 & 10 \\
  \hline
  8--12 Female  & 66  & 77  & 42  & 84  & 6  & 6  & 6  & 6  \\
  \hline
  8--12 Male    & 91  & 113 & 78  & 114 & 7  & 8  & 8  & 8  \\
  \hline
  12--18 Female & 56  & 65  & 42  & 63  & 6  & 6  & 6  & 6  \\
  \hline
  12--18 Male   & 92  & 96  & 73  & 96  & 7  & 7  & 7  & 8  \\
  \hline
\end{tabular}
\label{table:age_gender_groups}
\end{center}
\end{table}

\subsubsection{Maximum Subjects Matching Matching}

In this type of matching, we attempt to use the same proportion of desaturated
and nondesaturated patients in the train, test, and validation sets, while
maximizing the number of patients that we use for training our model. For
example, in the 5--8 Female age/gender group for NREM1 sleep, we have 75
subjects labeled as desaturated and 6 subjects labeled as undesaturated. We
select 25 desaturated subjects each for the train, test, and validation set. We
then divide the 6 undesaturated subjects into 2 subjects each for the train,
test, and validation sets. This method of age/gender matching allows us to
maximize the amount of data we use in our model training and testing pipeline.

\subsubsection{Equal Subjects Matching}\label{subsub:eq_match}

In this type of matching, we ensure that the number of desaturated and
nondesaturated patients in each group is equal in order to obtain the least
biased model possible. For example, in the 5--8 Female age/gender group for
NREM1 Sleep, we have 75 subjects labeled as desaturated and 6 subjects labeled
as undesaturated. In order to ensure that we have an equal number of subjects
in the desaturated and undesaturated groups for the train/test/validation splits
of our machine learning model, so we randomly split our 6 undesaturated group
subjects into groups of 2 for the train, test, and validation sets. We then
randomly select 6 subjects from the desaturated group to match the undesaturated
group, and divide them among the train, test, and validation set. We repeat this
process for each age and gender group. Since our results for this type of
matching are dependent on which subjects are randomly chosen from the
desaturated group, we repeat this type of matching 11 times and average model
results to obtain final performance results for our machine learning model.

\subsection{Machine Learning Methods}

After signal preprocessing we have a $7 \times 129 \times 61$ array for each EEG
epoch, with 7 representing the number of channels, 129 representing bins in the
frequency domain, and 61 representing bins in the time domain. We leverage a
ResNet based neural network as our machine learning model for
training~\cite{heDeepResidualLearning2016}. This allows us to treat each data
array as a 7-channel image, and apply learnable convolution filters to extract
features from the STFT data. All models are trained for 512 epochs using the Adam
optimizer and a learning rate of 1e-5~\cite{kingmaAdamMethodStochastic2017b}. We
additionally weight positive samples in our cross-entropy loss function using the inverse of
the frequency of occurrence of positive samples in the training set. We train on
our training set, and select a best model from our training process based on the
testing set. We then report results of our selected on a held-out validation set
to best characterize the generalization performance of the selected model. We
implement our machine learning pipeline in
PyTorch~\cite{paszkePyTorchImperativeStyle2019b}.

While the number of subjects in our dataset is balanced by our age/gender
matching schemes, our dataset an unequal number of positive and negative labels.
To accurately characterize the performance of our models, we leverage balanced
accuracy (BA) and area under receiver operating characteristic curve (AUC) as
our evaluation
metrics~\cite{brodersenBalancedAccuracyIts2010,fawcettIntroductionROCAnalysis2006}.
Balanced accuracy is calculated as the average of the true positive rate and
true negative rate of the classifier, and is robust to class imbalance. To
identify which samples the neural network model identifies as positive or
negative, we threshold the outputs of our neural network with a value of 0.5. To
calculate AUC, we plot the true positive rate and false positive rate for
various thresholds, and calculate the area under the resulting curve. We
leverage Python and scikit-learn to implement these
metrics~\cite{pedregosaScikitlearnMachineLearning2011a}.

\section{Results and Discussion}

When reporting results for all experiments, we write the number of subjects as
\# Train/\# Test/\# Validation, and write the validation set BA and validation
set AUC as the average value, followed by (Lower 95\% confidence interval, Upper
95\% confidence interval) for the balanced matching experiments since this
experiment is run 11 times. For the imbalanced matching experiments, we only
report the final validation accuracy, as the experiment is run once. Note that,
in maximum subjects matching, for each stage of sleep we use the same subjects
in each experiment. This allows for the closest comparison of results across
experiment types. In equal subjects matching, for each stage of sleep we use the
same subjects in each iteration (i.e., iteration 0 of each experiment uses the
same subjects, iteration 1 uses the same subjects, etc.). This allows for the
closest comparison of results across experiment types, and attempts to eliminate
any randomness caused by downselecting subjects during this type of matching.

\subsection{Classification of Oxygen Desaturations Across Patients}\label{section:extract}

The objective of this experiment is to understand if there is a potential
biomarker of oxygen desaturation in EEG data which persists across patients.
This experiment is a two-class problem. We label EEG data which is taken during
desaturation events from subjects who experience desaturations 1, and label EEG
data which is taken during normal sleep from subjects who do not experience
desaturations 0. To enforce data size consistency, we use the entire 30-second
EEG epoch, even if the desaturation is only part of the epoch. Results are
included in Table~\ref{table:imbalanced_extract_results} and
Table~\ref{table:balanced_extract_results}.

\begin{table}[h]
\caption{Cross-Patient Experiment Results -- Maximum Subjects Matching}
\centering
\resizebox{\columnwidth}{!}{%
  \begin{tabular}{|c|c|c|c|}
    \hline
    \textbf{Sleep Type} & \textbf{\# Subjects} & \textbf{Validation BA} & \textbf{Validation AUC} \\
    \hline
    \textbf{NREM1} & 329/329/329 & 0.760 & 	0.905 \\
    \hline
    \textbf{NREM2} & 397/397/397 & 0.625 & 0.700 \\
    \hline
    \textbf{NREM3} & 300/300/300 & 0.758 & 0.814 \\
    \hline
    \textbf{REM} & 429/429/429 & 0.531 & 0.648 \\
    \hline
  \end{tabular}
\label{table:imbalanced_extract_results}
}
\end{table}

\begin{table}[h]
\caption{Cross-Patient Experiment Results -- Equal Subjects Matching}
\centering
\resizebox{\columnwidth}{!}{%
  \begin{tabular}{|c|c|c|c|}
    \hline
    \textbf{Sleep Type} & \textbf{\# Subjects} & \textbf{Validation BA} & \textbf{Validation AUC} \\
    \hline
    \textbf{NREM1} & 28/28/28 & 0.732 (0.680, 0.784) & 0.799 (0.745, 0.852) \\
    \hline
    \textbf{NREM2} & 30/30/30 & 0.655 (0.602, 0.709) & 0.725 (0.657, 0.793) \\
    \hline
    \textbf{NREM3} & 30/30/30 & 0.763 (0.720, 0.807)& 0.830 (0.780, 0.879) \\
    \hline
    \textbf{REM} & 30/30/30 & 0.702 (0.637, 0.767) & 0.719 (0.649, 0.790) \\
    \hline
  \end{tabular}
  \label{table:balanced_extract_results}
}
\end{table}

From this experiment, we find that the model performs well for NREM1 (BA =
0.760), NREM2 (BA = 0.625), and NREM3 sleep (BA = 0.758), and not as well for
REM sleep (BA = 0.531), specifically in the maximum subjects matching case. The
maximum subjects and equal subjects matching results are within 95\% confidence
interval of each other for all sleep stages but REM. In general, oxygen
desaturation is lower during REM sleep; furthermore, if hypoventilation occurs,
oxygen desaturation can increase further than in other sleep
stages~\cite{choiSeveritySleepDisordered2016}. This may contribute to the poor
performance of the model in REM sleep during the maximum matching case, as
there are many examples of oxygen desaturation in REM sleep, but very few
examples of undesaturated REM sleep from healthy subjects, who already have
decreased oxygen saturation.

\subsection{Classification of Oxygen Desaturations Within Patients}\label{section:apnea}

The objective of this experiment is to understand if there is a potential
biomarker of oxygen desaturation in EEG data which can be identified only from
patients who experience sleep oxygen desaturations.  When selecting data, we use
the entire EEG epoch, even if the desaturation is only part of the epoch. This
experiment is a two-class problem. If the patient has desaturations, we include
all EEG epochs, labeling those that contain desaturations 1 and those that do
not contain desaturations 0. If the patient does not have desaturations, we do
not include any of their data. The subject counts are lower than in the
experiment described in Section~\ref{section:extract} since we only use the
subjects in the desaturated group. Results are included in
Table~\ref{table:imbalanced_apnea_results} and
Table~\ref{table:balanced_apnea_results}.

\begin{table}[h]
\caption{Within Patient Experiment Results -- Maximum Subjects Matching}
\centering
\resizebox{\columnwidth}{!}{%
  \begin{tabular}{|c|c|c|c|}
    \hline
    \textbf{Sleep Type} & \textbf{\# Subjects} & \textbf{Validation BA} & \textbf{Validation AUC} \\
    \hline
    \textbf{NREM1} & 315/315/315 & 0.639 & 0.675 \\
    \hline
    \textbf{NREM2} & 382/382/382 & 0.619 & 0.676 \\
    \hline
    \textbf{NREM3} & 285/285/285 & 0.705 & 0.794 \\
    \hline
    \textbf{REM}   & 414/414/414 & 0.583 & 0.609 \\
    \hline
  \end{tabular}
\label{table:imbalanced_apnea_results}
}
\end{table}

\begin{table}[h]
\caption{Within Patient Experiment Results -- Equal Subjects Matching}
\centering
\resizebox{\columnwidth}{!}{%
  \begin{tabular}{|c|c|c|c|}
    \hline
    \textbf{Sleep Type} & \textbf{\# Subjects} & \textbf{Validation BA} & \textbf{Validation AUC} \\
    \hline
    \textbf{NREM1} & 14/14/14 & 0.624 (0.587, 0.661) & 0.685 (0.639, 0.731) \\
    \hline
    \textbf{NREM2} & 15/15/15 & 0.597 (0.525, 0.669) & 0.642 (0.557, 0.728) \\
    \hline
    \textbf{NREM3} & 15/15/15 & 0.691 (0.629, 0.753) & 0.745 (0.665, 0.825) \\
    \hline
    \textbf{REM}   & 15/15/15 & 0.587 (0.541, 0.633) & 0.602 (0.532, 0.672) \\
    \hline
  \end{tabular}
\label{table:balanced_apnea_results}
}
\end{table}

We observe best performance in NREM3 sleep (BA = 0.705), and worst performance
in REM sleep (BA = 0.583), similarly to the experiment described in
Section~\ref{section:extract}. From this experiment, we conclude that we can
identify oxygen desaturations in EEG signals from the same subject with
reasonable accuracy in NREM1, NREM2, and NREM3 sleep; however, performance in
REM sleep is still poor. Since oxygen desaturation is naturally lower during REM
sleep, the model is not able to differentiate significant desaturations from
undesaturated data in single patients due to the frequent desaturations
occurring during sleep~\cite{choiSeveritySleepDisordered2016}. The undesaturated
data from subjects who experience desaturations occurs in conjunction with
significant desaturations during the same stage of sleep, leading to potential
features being mixed between the desaturated and undesaturated EEG epochs,
leading to poor model performance. We furthermore see that except for REM sleep
for the maximum matching experiment, performance is worse than in the
cross-patient experiment; this indicates that there is potentially some
fundamental difference between the EEG data of subjects who experience
desaturations and subjects who do not experience
desaturations.

\subsection{Identification of Latent Markers of Oxygen Desaturation}\label{section:whole}

The objective of this experiment is to understand if there is a potential latent
biomarker of oxygen desaturation contained in EEG data which persists when
oxygen desaturations are not occurring. This experiment is a two-class problem.
If the patient has desaturations, we identify EEG epochs that do not contain
desaturations, and label those 1. We eliminate the entirety of any EEG epochs
which contain desaturations. If the patient does not have desaturations, we
include all data for that sleep stage, and label it 0. Since some subjects
contain frequent apnea events and do not have any undesaturated data, subject
counts for this experiment vary from the subject counts described in
Section~\ref{section:extract}. Results are included in
Table~\ref{table:imbalanced_whole_results} and
Table~\ref{table:balanced_whole_results}.

\begin{table}[h]
\caption{Latent Marker Experiment Results -- Maximum Subjects Matching}
\centering
\resizebox{\columnwidth}{!}{%
  \begin{tabular}{|c|c|c|c|}
    \hline
    \textbf{Sleep Type} & \textbf{\# Subjects} & \textbf{Validation BA} & \textbf{Validation AUC} \\
    \hline
    \textbf{NREM1} & 199/185/168 & 0.550 & 0.733	\\
    \hline
    \textbf{NREM2} & 251/254/265 & 0.556 & 0.672 \\
    \hline
    \textbf{NREM3} & 171/167/167 & 0.645 & 0.799 \\
    \hline
    \textbf{REM}   & 284/290/277 & 0.644 & 0.691 \\
    \hline
  \end{tabular}
\label{table:imbalanced_whole_results}
}
\end{table}

\begin{table}[h]
\caption{Latent Marker Experiment Results -- Equal Subjects Matching}
\centering
\resizebox{\columnwidth}{!}{%
  \begin{tabular}{|c|c|c|c|}
    \hline
    \textbf{Sleep Type} & \textbf{\# Subjects} & \textbf{Validation BA} & \textbf{Validation AUC} \\
    \hline
    \textbf{NREM1} & 21/19/24 & 0.685 (0.632, 0.737) & 0.784 (0.743, 0.826) \\
    \hline
    \textbf{NREM2} & 24/23/25 & 0.572 (0.535, 0.608) & 0.585 (0.544, 0.627) \\
    \hline
    \textbf{NREM3} & 21/25/24 & 0.601 (0.523, 0.680) & 0.638 (0.538, 0.737) \\
    \hline
    \textbf{REM}   & 23/23/21 & 0.591 (0.551, 0.631) & 0.650 (0.598, 0.703) \\
    \hline
  \end{tabular}
\label{table:balanced_whole_results}
}
\end{table}

In NREM3 and REM sleep, our imbalanced matching shows the potential for a latent
biomarker with validation BA values of 0.645 and 0.644 respectively. The
difference between the balanced and imbalanced matching experiments is also
significant, indicating that using the imbalanced dataset induces some amount of
bias which leads to higher model performance in NREM3 and REM sleep. However,
this experiment results in comparable performance to the results from the within
patient oxygen desaturation experiment, indicating the potential for a
brain-based biomarker of oxygen desaturation.

\section{Conclusions}

In this work, we explore the ability of machine learning models to discriminate
between EEG signals taken during oxygen desaturation events during different
sleep stages and EEG signals taken during normal sleep. We find that these
models are able to identify EEG signals taken during oxygen desaturations with
reasonable accuracy, indicating the potential for a brain-based biomarker for
oxygen desaturations. Further work is required to isolate specific morphological
features in each EEG signal which indicate oxygen desaturation.

\bibliographystyle{IEEEtran}
\bibliography{smanjunath_arxiv_version}

\end{document}